# Spin-dependent transport in the p-type ultra-narrow silicon quantum well

N T Bagraev[1], M I Bovt[2], W Gehlhoff[3], O N Guimbitskaya[2], L E Klyachkin[1], A M Malyarenko[1], V A Mashkov[2], V V Romanov[2] and S A Rykov[2]

[1] Ioffe Physico-Technical Institute RAS, 194021, St.Petersburg, Russia
[2] St.Petersburg Polytechnical University, St.Petersburg, 195251, Russia
[3] Technische Universitaet Berlin, D-10623 Berlin, Germany

E-mail: impurity.dipole@mail.ioffe.ru

**Abstract**. We present the findings of the spin-dependent transport of the 2D holes in the p-type ultra-narrow self-assembled silicon quantum well (Si-QW) confined by the superconductor barriers on the n-type Si (100) surface. The Aharonov-Casher conductance oscillations caused by the Rashba spin-orbit interaction are found to interplay with the quantization phenomena induced by the creation of the two-dimensional subbands in Si-QW and the quantum subbands due to the Andreev reflection from the superconductor barriers

Semiconductor silicon is well known to be the principal material for micro - and nanoelectronics. Specifically, the developments of the silicon planar technology are a basis of the metal-oxygen-silicon (MOS) structures and silicon-germanium (Si-Ge) heterojunctions that are successfully used as elements of modern processors [1]. Just the same goals of future high frequency processors especially to resolve the problem of quantum computing are proposed to need the application of the superconductor nanostructures that represent the Josephson junction series [2]. Therefore the manufacture of superconductor device structures in frameworks of the silicon planar technology seems to give rise to new generations in nanoelectronics. Furthermore, one of the best candidate on the role of the superconductor shells for silicon nanostructure appears to be the δ - barriers heavily doped with boron that confine the high mobility silicon quantum wells (Si-QW) of the p-type located on the n-type Si (100) surface [3]. The findings of the electrical resistivity, thermo-emf and magnetic susceptibility measurements are actually evidence of the superconductor properties for the δ - barriers heavily doped with boron, which are revealed at high density of holes in the Si-QW. These silicon nanostructures embedded in superconductor shells are shown to be type II high temperature superconductors (HTS) with $T_c$=145 K and $H_{c2}$=0.22 T to be very promising as the sources and recorders of the THz irradiation by controlling the vortex transport phenomena [4].

The superconductor silicon nanostructures are self-assembled on the n-type Si (100) wafers during preliminary oxidation and subsequent short-time diffusion of boron by the CVD method [3]. The preparation of oxide overlayers on silicon monocrystalline surfaces is known to be favourable to the generation of the excess fluxes of self-interstitials and vacancies that exhibit the predominant crystallographic orientation along a <111> and <100> axis, respectively (figure 1a) [3,5]. In the initial stage of the oxidation, thin oxide overlayer produces excess self-interstitials that are able to create small microdefects, whereas oppositely directed fluxes of vacancies give rise to their annihilation (figures 1a and 1b). Since the points of outgoing self-interstitials and incoming vacancies appear to be defined by the positive and negative charge states of the reconstructed silicon dangling bond [3,5], the dimensions of small microdefects of the self-interstitials type near the Si (100) surface have to be restricted to 2 nm. Therefore, the distribution of the microdefects created at the initial stage of the oxidation seems to represent the fractal of the Sierpinski Gasket type with the built-in self-assembled Si-QW (figure 1b). Although Si-QWs embedded in the fractal system of self-assembled microdefects are of interest to be used as a basis of optically and electrically active microcavities in optoelectronics and nanoelectronics, the presence of dangling bonds at the interfaces prevents such an application. Therefore, subsequent short-time diffusion of boron would be appropriate for the passivation of dangling bonds and other defects created during previous oxidation of the Si (100) surface thereby assisting the transformation of the arrays of microdefects in the neutral δ - barriers confining the ultra-narrow, 2nm, Si-QW (figure 2a).
We have prepared the p-type self-assembled Si-QWs with different density of holes ($10^9 \div 10^{12}$ cm$^{-2}$) on the Si (100) wafers of the n-type in frameworks of the conception discussed above and identified the properties of the two-dimensional high mobility gas of holes by the cyclotron resonance (CR), Hall-effect and infrared Fourier and tunneling spectroscopy methods. Besides, the secondary ion mass spectroscopy (SIMS) and scanning tunneling microscopy (STM) studies have shown that the δ - barriers, 3 nm, heavily doped with boron, $5 \cdot 10^{21}$ cm$^{-3}$, represent really alternating arrays of undoped and doped tetrahedral dots with dimensions restricted to 2 nm. The value of the boron concentration determined by the SIMS method seems to indicate that each doped dot located between undoped dots contains two impurity atoms of boron. The EPR and the thermo-emf studies show that these boron pairs are the trigonal dipole centres, $B^+$-$B^-$, which are caused by the negative-U reconstruction of the shallow boron acceptors, $2B_0 => B^+ + B^-$. In common with the other solids that consist of small bipolarons, the delta barriers containing the dipole boron centres have been found to be in an excitonic insulator regime at the density of holes in the Si-QW lower than $10^{11}$ cm$^{-2}$. However, the electrical resistivity, thermo-emf and magnetic susceptibility measurements demonstrate that the high density of holes in the Si-QW ($>10^{11}$ cm$^{-2}$) gives rise to the superconductor properties for the δ – barriers in frameworks of the mechanism of the single-hole tunnelling through the negative-U centres [4].

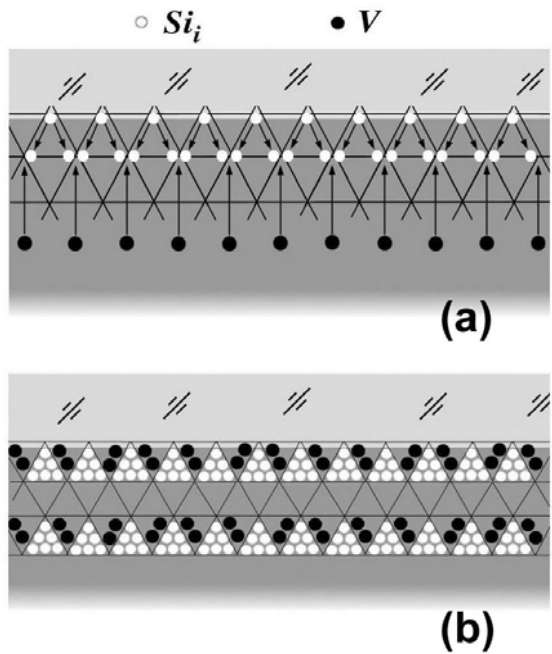
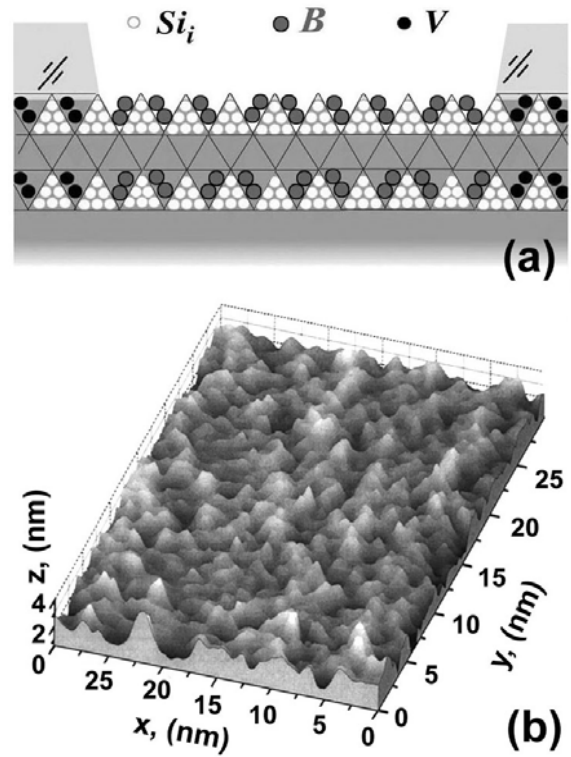

**Figure 1.** A scheme of self-assembled silicon quantum wells (Si-QW) confined by the δ - barriers on the Si (100) surface.
(a) The white and black balls label the self-interstitials and vacancies forming the excess fluxes oriented crystallographically along a <111> and <100> axis that are transformed to small microdefects.
(b) The silicon quantum wells between the layers of microdefects that are produced by performing thin oxide overlayer.

**Figure 2.** (a) In the process of short-time diffusion **t**he atoms of boron replace the positions of vacancies thereby passivating the layers of microdefects and forming the neutral δ - barriers.
(b) Scanning tunneling microscopy image of the top δ - barrier heavily doped with boron that confines the p-type Si-QW prepared on the n-type Si (100) surface. Light tetrahedrons result from the microdefects of the self-interstitials type. X∥[001], Y∥[010], Z∥[100].

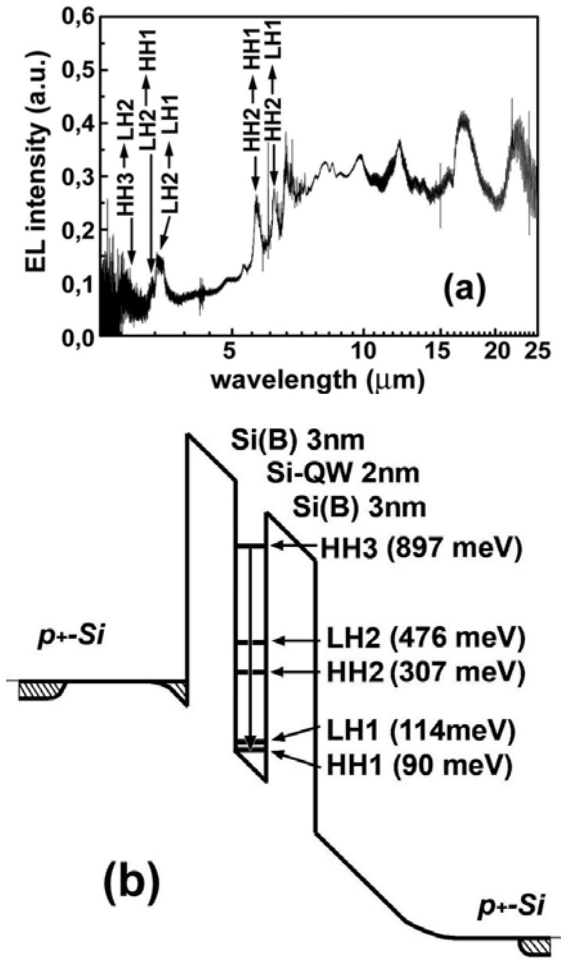
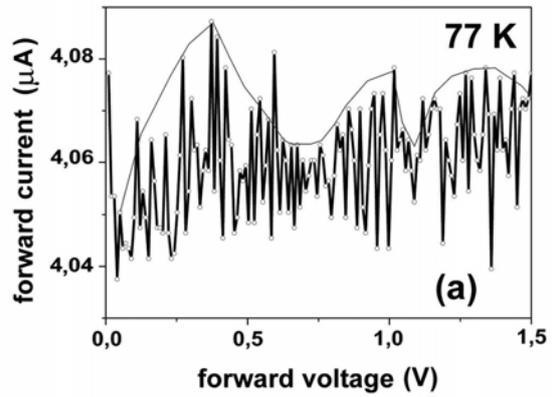

**Figure 3.** (a) Electroluminescence spectrum caused by optical transitions between two-dimensional subbands of holes in the p-type Si-SQW confined by the HTS barriers.
(b) Two-dimensional subbands of holes in self-assembled silicon quantum well.

**Figure 4.** (a) Forward CV characteristic that identifies the energy positions of the two-dimensional subbands of 2D holes in the p-type Si-QW shown as the spectral maxima as well as mesoscopic current oscillations, which define quantum subbands caused by the Andreev reflection from the HTS barriers (b).

The energy positions of two-dimensional subbands for the light and heavy holes in the Si-QW were determined by studying the far-infrared electroluminescence spectra obtained at T=300 K with the infrared Fourier spectrometer IFS-115 Brucker Physik AG (figures 3a and 3b). The results analyzed are in a good agreement with corresponding calculations following by Ref [6] if the width of the Si-QW, 2nm, is taken into account. Besides, the two-dimensional subbands of the 2D holes are revealed by the forward CV characteristic that exhibits also the spectrum of the tunnelling current due to the Andreev refelection from the HTS barriers confining Si-QW (figure 4a). This tunnelling spectrum gives rise to the formation of the additional quantum subbands of the 2D holes (see figure 4b) that establish the spectral range of the THz generation. Specifically, several modes of the THz irradiation in the interval from 0.12 THz to 5.3 THz are found under the gate voltage applied perpendicularly to the Si-QW plane (figure 3a). Moreover, the THz generation that is attended with the mesoscopic curent fluctuations appears to result in the high mobility of the 2D hole gas, which is revealed by the Hall measurements $2.7 \times 10^5$ cm$^2$/Vs at T = 77K ($p_{2D}$ = $4 \times 10^{11}$ cm$^{-2}$), and to favour the high-frequency analogue of the Shubnikov oscillations [7] (see figures 5 and 6).

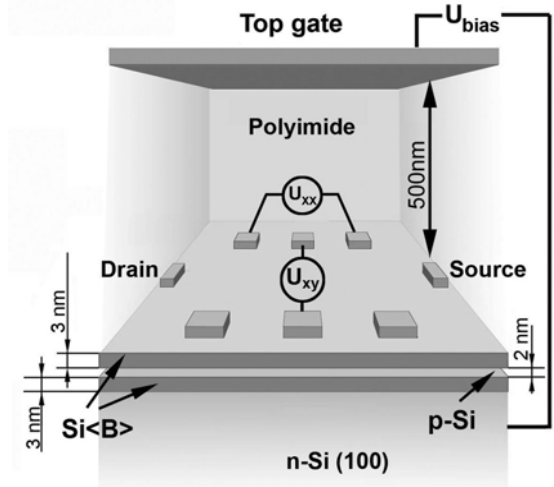 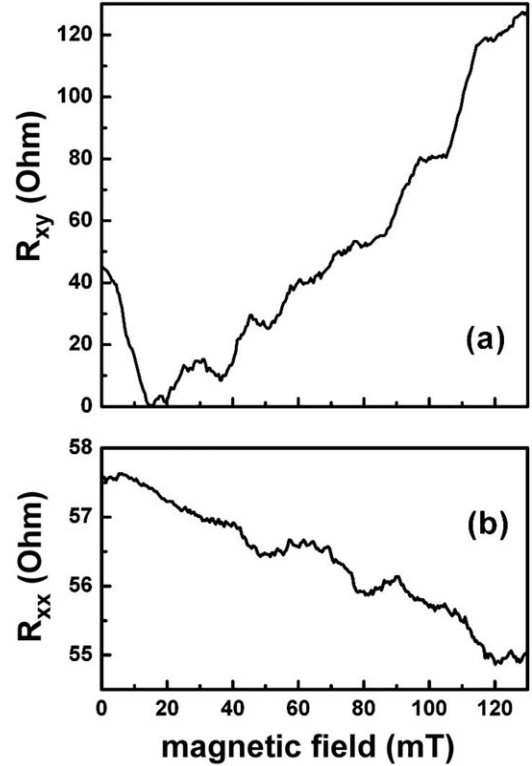

**Figure 5.** The scheme of the p-type Si-QW that is prepared in frameworks of the Hall geometry on the n-type Si (100) surface, which is provided by the top gate to vary the value of the Rashba spin-orbit interaction.

**Figure 6.** $R_{xx}$ and $R_{xy}$ vs magnetic field value that is applied perpendicularly to the plane of the p-type Si-QW on the n-type Si (100) surface ($I_{ds}$=10 nA).

The interplay of the Andreev quantization processes and the Rashba spin-orbit interaction (SOI) appear to result in the spin polarization of the 2D holes in the transport processes by varying the top gate voltage (see figure 5). Thus, the observation of the spin transistor effect becomes it possible without the injection of the spin-polarized carriers from the iron contacts as proposed in the classical version of this device [8]. The Aharonov-Casher (AC) conductance oscillations found by varying the top gate voltage that controls the Rashba SOI value seems to verify the principal characteristic of the spin transistor (figure 7). The modulation of the conductance shown in figure 7 appears to result from the phase shift induced by the effective magnetic field, which is created by the Rashba SOI and represents the vector product of the external electric field and the carrier wavevector that is directed in the Si-QW plane perpendicularly to the movement of the carriers [8,9],

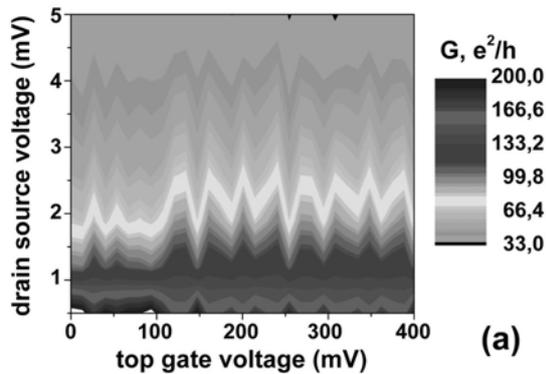 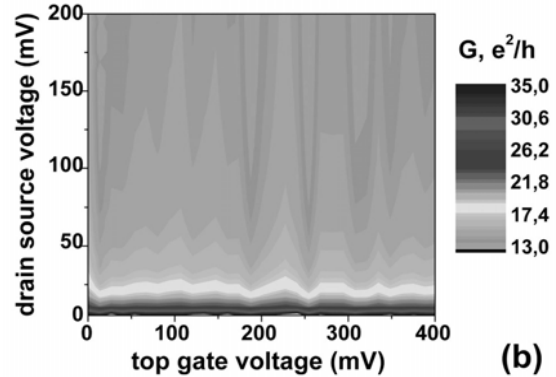

**Figure 7.** The AC conductance oscillations observed in the studies of the p-type Si-QW on the n-type Si (100) surface that exhibit the spin transistor effect at T=77 K by varying the top gate voltage, which controls the value of the Rashba SOI.

**Figure 8.** The quenching of the AC conductance oscillations by increasing the drain-source voltage that gives rise to both the drop of the spin-lattice relaxation time and the transitions between different subbands of holes.

$$\mathbf{B}_{eff} = \frac{\alpha}{g_B \mu_B}[\mathbf{k} \times \mathbf{e}_z] \qquad (1)$$

where $\mu_B$ is the Bohr magneton and $\alpha$ is the Rashba SOI parameter.

These conductance oscillations appear to smooth over increasing the drain-source voltage that gives rise to both the drop of the spin-lattice relaxation time and the transitions between different subbands of 2D holes (figure 8). The period of the conductance oscillations is dependent on the value of the effective field induced by the Rashba SOI, which is determined by the Si-QW characteristics and the effective mass value of the 2D holes [9]. Therefore the findings shown in figure 7 are advantadgeously used to define the effective mass of the 2D holes. The magnitude derived from the value of the period of the AC conductance oscillations, $6.7 \times 10^{-3}$ $m_0$, is in a good agreement with data obtained from the angular dependences of the cyclotron resonance and the Hall measurements (figure 6). It should be noted that the low effective mass of the 2D holes that seems to be caused by the interplay of the Andreev quantization and the Rashba SOI allows the findings of the spin transistor effect at T=77K.

In conclusions, the best compromise on the presence of the $p^+$-n junction revealed by the CV characteristics and the control of the Rashba spin-orbit interaction by the gate-voltage applied to the HTS – Si-QW sandwich opens up new possibilities for the spin-dependent transport, even though the external magnetic field is absent.

This work was supported by SNSF in frameworks of the programme "Scientific Cooperation between Eastern Europe and Switzerland, Grant IB7320-110970/1.